\documentclass{sig-alternate-10}
\usepackage[english]{babel}
\usepackage{hyperref}
\usepackage{amsmath, amsfonts,xfrac,amssymb,fancyhdr}

\usepackage{mathtools}

\usepackage{setspace}

\usepackage{colortbl}

\usepackage{fixltx2e, ifthen, sidecap, wrapfig, xspace}

\usepackage{mathrsfs, stmaryrd}

\usepackage{
 calc, enumitem, makeidx,bbold
}

\usepackage[font=small,format=plain,labelfont=sc,textfont=it]{caption}
\usepackage[T1]{fontenc}
\usepackage[utf8]{inputenc}
\usepackage[all]{xy}
\usepackage[geometry]{ifsym}
\usepackage[cmyk]{xcolor}

\newcommand{\Nat}{\mathbb N}

\renewcommand{\k}{\mathbb{K}}

\newcommand{\str}[1]{\mathbb{#1}}
\newcommand{\RR}{\str R}
\newcommand{\eps}{\varepsilon}
\renewcommand{\Pr}[1]{\mathbb P[#1]}

\renewcommand{\subset}{\subseteq}
\DeclareMathOperator{\codim}{codim}
\DeclareMathOperator{\vc}{VC_{\k}}
\DeclareMathOperator{\vcs}{VC^\ast_{\k}}
\DeclareMathOperator{\gc}{GC}
\DeclareMathOperator{\agm}{AGM}

\begin{document}

\conferenceinfo{PODS'16,} {June 26-- July 01, 2016, San Francisco, USA.}
\CopyrightYear{2016}
\crdata{.}
\clubpenalty=10000
\widowpenalty = 10000

\title{Entropy bounds for conjunctive queries\\
with functional dependencies 
}

\numberofauthors{2} 
\author{
\alignauthor
Tomasz Gogacz\\
       \affaddr{School of Informatics}\\
       \affaddr{University of Edinburgh}\\
       \affaddr{Edinburgh, UK}\\
       \email{tgogacz@inf.ed.ac.uk}
\alignauthor
Szymon Toruńczyk\\
       \affaddr{Institute of Computer Science}\\
       \affaddr{University of Warsaw}\\
       \affaddr{Poland}\\
       \email{szymtor@mimuw.edu.pl}
}

\date{\today}

	\maketitle

	\begin{abstract}
We study the problem of finding the worst-case bound for the size of the result $Q(\str D)$ of a fixed conjunctive query $Q$ applied to a database $\str D$ satisfying given functional dependencies. We provide a precise characterization of this bound in terms of entropy vectors, 
and in terms of finite groups. In particular, we show that an upper bound provided by Gottlob, Lee, Valiant and Valiant~\cite{gottlob} is tight, answering a question from their paper. Our result generalizes the bound due to Atserias, Grohe and Marx~\cite{agm}, who consider the case without functional dependencies. Our result shows that the problem of computing the worst-case size bound, in the general case, is closely related to difficult problems from information theory.
	\end{abstract}
\noindent {Query languages}




\newdef{definition}{Definition}
\newdef{example}{Example}
\newdef{remark}{Remark}
\newtheorem{conjecture}{Conjecture}
\newtheorem{theorem}{Theorem}
\newtheorem{lemma}{Lemma}
\newtheorem{corollary}{Corollary}
\newtheorem{prop}{Proposition}
\newtheorem{claim}{Claim}
\newtheorem{fact}{Fact}

\newcommand{\var}{{\bf V}}
\newcommand\set[1]{\ensuremath{\{#1\}}}

\newcommand{\Ff}{{\mathcal F}}
\newcommand{\Vv}{\ensuremath{\mathcal V}\xspace}
\newcommand{\Gg}{{\mathcal G}}
\newcommand{\im}[1]{\textrm{Im}(#1)}

\newcommand{\R}{\mathbb{R}}

\section{Introduction}\label{sec:intro}
Given a natural join query $Q$ we would like to determine a bound $\alpha\in\RR$ 
such that for every  database~$\str D$, the inequality 
\begin{align}\label{eq:main}
|Q(\str D)|\ \le\  c\cdot |\str D|^{\alpha}
\end{align}
holds, for some multiplicative factor $c$ depending on $Q$. Here, $|\str D|$ denotes the  size of the largest table in the database $\str D$, and $|Q(\str D)|$ denotes the size of the result of the query applied to $\str D$. 
Above, we consider all databases $\str D$ over a fixed schema, containing the relation names which appear in~$Q$. In the general problem, which is the main focus of this paper, we may additionally impose some functional dependencies, and require that $\str D$ satisfies them.

Obviously, in~\eqref{eq:main} we can always take $\alpha=|Q|$, the number of relation names appearing in the query~$Q$. Define $\alpha(Q)$  as the infimum of all values $\alpha$ for which there exists a multiplicative factor $c$ so that~\eqref{eq:main} holds for all databases $\str D$.
For example, if $Q_1(x,y,z)=R(x,y)\land S(y,z)$ then 
it is not difficult to see that $\alpha(Q_1)=2$. Indeed, $$|Q_1(\str D)|\le |R(\str D)|\cdot |S(\str D)|\le |\str D|^2,$$ and conversely, one can construct a database $\str D$ with 
\begin{align*}
R(\str D)&=X\times Y,\\
S(\str D)&=Y\times Z,
\end{align*}
for some finite sets $X,Z$ of arbitrarily large size $N$ and $Y$ of size $1$, and then $|Q_1(\str D)|=|X\times Y\times Z|=N^2$, whereas $|R(\str D)|=|S(\str D)|=N$.
Now consider $Q_2(x,y,z)=R(x,y)\land S(y,z)\land T(z,x)$. The trivial bound gives $\alpha(Q_2)\le 3$. However, since $Q_2(\str D)\subseteq Q_1(\str D)$ for every $\str D$, it follows immediately that $\alpha(Q_2)\le 2$. With some effort, one can show that in the absence of functional dependencies, $\alpha(Q_2)=3/2$. This is a consequence of a more general result due to Atserias, Grohe and Marx~\cite{agm}, which we recall now.

Consider the hypergraph  whose vertices are the variables appearing in $Q$, and for each relation name $R$ in $Q$ there is a hyperedge containing those variables which appear in $R$. A \emph{fractional edge covering} of a hypergraph assigns a positive rational number to each of its hyperedges, so that for every vertex,  the numbers assigned to the adjacent hyperedges 
sum up to at least $1$. The total weight of a fractional edge covering is the sum of the numbers assigned to all the hyperedges.
Define $\agm(Q)$ as the least possible total weight of 
a fractional vertex covering of the hypergraph associated to $Q$. The AGM bound then states that $\agm(Q)=\alpha(Q)$,
in the absence of functional dependencies.
For example, the hypergraph obtained from the query $Q_2$ is the triangle, and assigning $1/2$ to each edge gives a fractional edge covering with total weight $3/2$, which is the least possible. 
Hence, $\alpha(Q_2)=\agm(Q_2)=3/2$.

Fractional edge coverings of a hypergraph can be computed efficiently using linear programming. So the problem of finding $\alpha(Q)$ is solved, when there are no functional dependencies. 
However, if we consider only those databases which satisfy a given set of functional dependencies, the problem of computing $\alpha(Q)$ remains unsolved. (To see that the value of $\alpha(Q)$ may change in the presence of functional dependencies, observe that $\alpha(Q_1)=1$ assuming that $y$ is a key in $R$ and in $S$.)

In this paper, we make progress towards characterizing the value $\alpha(Q)$ in the presence of functional dependencies. In particular, we show that $\alpha(Q)$ can be characterized in two ways: as an entropy bound $H(Q)$, and in terms of a number $\gc(Q)$ derived from systems of finite groups. 
The bound $\alpha(Q)\le H(Q)$ was observed in~\cite{gottlob},
and it was left as an open problem whether equality holds for all
queries~$Q$, in the presence of functional dependencies.
We answer this question affirmatively, 
by providing a matching lower bound based on a construction using finite groups.
However, we do not know how to effectively compute the bound~$\alpha(Q)$.
Moreover, our results demonstrate that this problem is closely connected to notorious problems from information theory.

Finally, we discuss how to treat general conjunctive queries (under the usual semantics and under the bag semantics), and show some preliminary results concerning the computation of the result $Q(\str D)$.
\section{Main result and consequences}\label{sec:results}
The main result of this paper, Theorem~\ref{thm:optimization}, expresses $\alpha(Q)$
in terms of entropy.
Below we also state a more general result, Theorem~\ref{thm:simple-stat}. 
We start from recalling the notion of the entropy cone 
from information theory below;
for precise definitions and notation conventions,
 see Section~\ref{sec:prelim}. 
 After that, we formulate the main result of this paper.
 Next, we show how some known results follow from this result, or its generalization.
In Theorem~\ref{thm:groups} we show that in order to  achieve the worst-case size increase $\alpha(Q)$, it is enough to consider very symmetric databases.



%
%

\paragraph{Entropy cone}
For a random variable $V$, let $H(V)$ denote its entropy. All random variables in this paper assume finitely many values. All logarithms are in base $2$.

\newcommand{\ent}{\mathrm{ent}}
Fix a finite set $X$.  Let $U=(U_x)_{x\in X}$ be 
a family of random variables indexed by $X$. For $Y\subset X$, let  $U[Y]$ denote the joint  random variable $(U_y)_{y\in Y}$. The variable $U[Y]$ can be seen as a random variable taking as values tuples indexed by $Y$.
 Consider the real-valued vector $\ent(U)$,  indexed by subsets $Y$ of~$X$, such that
 $\ent(U)_Y=H(U[Y])$  for $Y \subset X$.
Vectors of the form $\ent(U)\in \mathbb R^{P(X)}$, where $U$ is a family of random variables indexed by $X$, are called \emph{entropy vectors} (or \emph{entropic vectors}) with ground set $X$~\cite{ZhangY98}.

 The set of all entropy vectors with ground set $X$ forms a subset of  $\mathbb R^{P(X)}$,
 denoted $\Gamma_X^*$. Its topological closure $\overline {\Gamma_X^*}$
  is a convex cone. The sets $\Gamma_X^*$ and $\overline {\Gamma_X^*}$ are well studied, however, to date, they lack effective descriptions when $|X|\ge 4$. It is known that the closed cone  $\overline {\Gamma_X^*}$ is a polyhedron if $|X|\le 3$, and is not a polyhedron if $|X|>3$ (i.e. it is not described by finitely many linear inequalities).
    Entropy vectors $v\in \Gamma_X^*$ (and hence, also all $v\in \overline{\Gamma_X^*}$) satisfy the submodularity property,
    expressing \emph{Shannon's inequality} for information:
  \begin{align}\label{eq:submodularity}
  v_{Y\cup Z}+v_{Y\cap Z}\le v_Y+v_Z\text{\quad for }Y,Z\subset X.
  \end{align}

\paragraph{Main result} 
For a relation name $R$ in a schema $\Sigma$, by $\var(R)$ we denote the set of attributes of $R$. For $X\subset \var(R)$ and $x\in\var(R),$ we write $R:X\mapsto x$ to denote the functional dependency (fd) requiring that in $R$, the values of attributes $X$ determine the value of the attribute $x$. The value $\alpha(Q)$ is defined as in the introduction, taking into account all databases $\str D$ over the schema $\Sigma$ which satisfy a given set of functional dependencies $\Ff$.
The main result of this paper is the following.

\begin{theorem}  \label{thm:optimization}
  Fix a schema $\Sigma$ and a set of functional dependencies $\cal F$.
  Let $Q$ be a natural join query with variables~$X$.
    Then  $\alpha(Q)$ is equal to the maximal value of $v_X$,
    for $v$ ranging over $\overline{\Gamma_X^*}$ and satisfying:
    $$\begin{cases}
      v_{\var(R)}\le 1&\text{ for }R\in Q,\\
      v_{Z\cup \set{z}}=v_Z &\text{ for every fd $R:Z\mapsto z$ in $\cal F$.}
    \end{cases}$$
\end{theorem}

We now show how some results known previously can be obtained as consequences of Theorem~\ref{thm:optimization}.

 Relaxing the condition that $v\in \overline{\Gamma_X^*}$
to the condition that~$v$ is submodular gives the following
upper bound on $\alpha(Q)$, which is
equivalent to a bound in~\cite{gottlob}.
\begin{corollary}
  \label{cor:submodular}
    Let $s(Q)$ be the maximal value of $v_X$,
    for $v$ ranging over $\mathbb R^{P(X)}$ and satisfying:
    $$\begin{cases}
      v_{Y\cup Z}+v_{Y\cap Z}\le v_Y+v_Z&\text{ for }Y,Z\subset X,\\
      v_{\var(R)}\le 1&\text{ for }R\in Q,\\
      v_{Z\cup \set{z}}=v_Z &\text{ for every fd~$R:Z\mapsto z$ in $\cal F$.}
    \end{cases}$$
    Then $\alpha(Q)\le s(Q)$.
\end{corollary}
Note that the bound $s(Q)$ can be computed by linear programming.

We show how the AGM bound can be deduced from Corollary~\ref{cor:submodular}. Assume that the set of functional
dependencies is empty. It is easy to see that then 
in the above linear program describing $s(Q)$, the maximum is achieved for a vector $v$ such that
$v_Y=\sum_{x\in Y} v_{\set{x}}$ for $Y\subset X$. Therefore, we get the following.

\begin{corollary}
  \label{cor:AGM-upper}
  In the absence of functional dependencies,
    $\alpha(Q)\le s(Q)$, where $s(Q)$ is the maximal value of $\sum_{x\in X} v_x$,
    for $v$ ranging over $\mathbb R^X$ and satisfying
      \begin{align}\label{eq:vertex-packing}
      \sum_{x\in\var(R)}v_x\le 1\text{\quad for }R\in Q.        
      \end{align}
\end{corollary}

\begin{proof}[of the AGM bound]
The linear program~\eqref{eq:vertex-packing} corresponds to computing the 
\emph{fractional vertex packing number}, and is dual 
to the linear program computing the fractional edge covering number. By strong duality, $s(Q)=\agm(Q)$, and therefore, $\alpha(Q)\le \agm(Q)$ by the above corollary.
The converse inequality, $\alpha(Q)\ge \agm(Q)$, is the easier part
of the AGM bound, and is shown by constructing a database
from a given fractional vertex packing, as follows.
Let $v=(p_x/q)_{x\in X}$ be a rational solution to the fractional vertex packing problem with common denominator $q\in \Nat$. For $Y\subset X$, let $p_Y$ denote
$\sum_{x\in Y}p_x$. In an optimal solution, we have $p_{X}/q=s(Q)$ and  $\max_{R\in Q}(p_{\var(R)}/q)=1$.

Choose an arbitrary integer $N>1$, and for each $x\in X$,
a set $V_x$ with $N^{p_x}$ elements.
Construct a database $\str D$ so that $R(\str D)=\prod_{x\in \var(R)}V_x$. It is easy to see that
$|R(\str D)|=N^{p_{\var(R)}}$ and
$|Q(\str D)|=N^{p_X}$.
In particular, $$\frac{\log|Q(\str D)|}{\log|\str D|}=\frac{p_X}{\max_{R\in Q}p_{\var(R)}}=s(Q).$$ Since $N$ 
can be taken arbitrarily large, this proves $\alpha(Q)\ge s(Q)$. Together with Corollary~\ref{cor:AGM-upper},
this shows that in the absence of functional dependencies, $s(Q)=\alpha(Q)=\agm(Q)$. 
\end{proof}

\paragraph{Symmetric databases}
The above proof of the AGM bound shows that in the absence of functional dependencies, the databases $\str D$
for which the size-increase $\frac{\log|Q(\str D)|}{\log|\str D|}$ achieves the bound $\alpha(Q)$ are of a very simple,
specific form: each table is a full Cartesian product.
It follows from~\cite{gottlob} that in the presence 
of functional dependencies, this is no longer the case:
databases of this form, satisfying the given functional dependencies, are arbitrarily far from reaching the value of $\alpha(Q)$.

In this paper, we improve the construction 
of worst-case databases in the presence of functional dependencies, by constructing databases
which are arbitrarily close to achieving the bound $\alpha(Q)$. Interestingly these databases have a very symmetric structure, and their construction uses finite groups.

For a finite group $G$
and a set $X$, recall that a (left) action of $G$ on $X$
is a mapping $G\times X\to X$ denoted $(g,x)\mapsto g\cdot x$, such that $(g\cdot h)\cdot x=g\cdot (h\cdot x)$
for $g,h\in G$ and $x\in X$. We say that the action is \emph{transitive} if for every $x,y\in X$
there is $g\in G$  such that $g\cdot x=y$. 
Transitive group actions are very special, and correspond (up to isomorphism) to subgroups of $G$, where a subgroup $H$ defines the action of $G$ on the coset space $G/H$
of left cosets. 

For a fixed group $G$, by a \emph{$G$-symmetric} database we mean a database $\str D$
together with an action of $G$ on the set of all values appearing in all tables of $\str D$,
such that the componentwise action of $G$
on each table $R(\str D)$ is transitive
(the componentwise action is given by
$(g\cdot r)[x]=g\cdot (r[x])$ for $g\in G,r\in R(\str D)$
and $x\in\var(R)$).
 A \emph{symmetric} database is a database $G$
which is $G$-symmetric for some finite group $G$.
For example, in the proof of the AGM bound presented above,
the constructed database $\str D$ is $G$-symmetric, for $G=\prod_{x\in\var(Q)} S_N^{p_x}$, where $S_N$
denotes the permutation group on $N$ elements. 

The second main result of this paper is the following.

\begin{theorem}\label{thm:groups}
  Fix a schema $\Sigma$, functional dependencies $\Ff$,
  and a natural join query $Q$.  Then, for each $\eps>0$ there are arbitrarily large symmetric databases~$\str D$ with $$\frac{\log|Q(\str D)|}{\log|\str D|}>\alpha(Q)-\eps.$$ \end{theorem}

Symmetric databases are essential in our proof of the lower bound given in Theorem~\ref{thm:optimization}.

\paragraph{Simple statistics}
We show how Theorem~\ref{thm:optimization} can be deduced from a more general result, which we now state.
For a database $\str D$ over the schema of $Q$,
its \emph{simple log-statistics} is the vector
$\mathrm{sls}(\str D)=\left({\log|R(\str D)|}\right)_{R\in Q}.$

\begin{theorem}\label{thm:simple-stat}
  Fix a schema $\Sigma$ and a set of functional dependencies $\cal F$.
    Let $Q$ be a natural join query with variables $X$,
  and let $s=(s_R)_{R\in Q}$ be a vector of nonnegative real numbers. 
  Let $\beta(Q,s)$ be
 the maximal value of $v_X$,
  for $v$ ranging over $\overline{\Gamma_X^*}$ and satisfying:
  $$\begin{cases}
    v_{\var(R)}\le  s_R&\text{ for }R\in Q,\\
    v_{Z\cup \set{z}}=v_Z &\text{ for every fd $R:Z\mapsto z$ in $\cal F$.}
  \end{cases}$$
    Then,   
  $$\sup_{\str D}{\log|Q(\str D)|}=\limsup_{\str D}{\log|Q(\str D)|}=\beta(Q,s),$$
  where $\str D$ ranges over all finite databases satisfying the  functional dependencies $\cal F$ and such that $\mathrm{sls}(\str D)\le s$
  componentwise.
\end{theorem}
Theorem~\ref{thm:optimization} is an immediate consequence of 
Theorem~\ref{thm:simple-stat}, obtained by setting $s_R=
{\log|\str D|}$ for each $R\in Q$,
and using the fact that $\overline{\Gamma_X^*}$ is a cone.
In this paper, we present in detail only the proof of Theorem~\ref{thm:optimization}. The proof of Theorem~\ref{thm:simple-stat} proceeds similarly, and will be presented in the full version of the paper.

We remark that Theorem~\ref{thm:simple-stat}
can be used to derive the following, more precise variant of the AGM bound.
\begin{corollary}[\cite{agm}]\label{cor:agm-full}
In the absence of functional dependencies, $|Q(\str D)|\le \prod_{R\in Q}
 |R(\str D)|^{w_R}$, where $(w_R)_{R\in Q}$ is any solution to the fractional edge covering problem.
\end{corollary}
Indeed, to deduce this, repeat the reasoning
used above when deriving Corollary~\ref{cor:AGM-upper}
from Theorem~\ref{thm:optimization}, by relaxing the condition $v\in \overline{\Gamma_X^*}$ 
in Theorem~\ref{thm:simple-stat} to submodularity of $v$.
We leave the details to the reader.

\paragraph{Geometric inequalities} 
As noted elsewhere~\cite{NgoPRR12,Friedgut04,BeameKS13}, Corollary~\ref{cor:agm-full} 
provides an upper bound on the size of a finite set in multi-dimensional space, in terms of the sizes of its projections. It implies the discrete versions of many inequalities from geometry and analysis, such as H\"older's inequality, Cauchy-Schwartz inequality, Loomis-Whitney inequality, Bollob\'as-Thomason inequality. 
As an illustration, we show how the Loomis-Whitney inequality can be derived from~Corollary~\ref{cor:agm-full}.

For $1\le j\le n$, let $\pi_j:\mathbb R^n\to \mathbb R^{n-1}$ denote the projection along the $j$th coordinate axis.
\begin{theorem}[Loomis-Whitney inequality~\cite{LW49}]
  \label{thm:lw}
\emph{Continuous variant:}  Let $E\subset \mathbb R^d$ be a measurable set.
  Then, for $\lambda_n$ denoting the $n$-dimensional Lebesgue measure,  
   $$\lambda_d(E)\le \prod_{j=1}^d\lambda_{d-1}(\pi_j(E))^{1/(d-1)}.$$
\emph{Discrete variant:}
  Let $E\subset \mathbb R^d$ be a finite set.
  Then $$|E|\le \prod_{j=1}^d|\pi_j(E)|^{1/(d-1)}.$$
\end{theorem}

The discrete variant of Theorem~\ref{thm:lw} can be seen as a consequence of Corollary~\ref{cor:agm-full}, as follows.
Consider a schema $\Sigma$ with relations $R_i$, for $i=1,\ldots,d$, and attributes $x_i$, for $i=1,\ldots,d$,
where $\var(R_i)=\set{x_j:j\neq i, 1\le j\le d}$.
For a finite set $E\subset \mathbb R^d$,
where $E$ is naturally viewed as a set of functions $r:\set{x_1,\ldots,x_d}\to\mathbb R$,
define a database~$\str D$ over $\Sigma$ with $R_i(\str D)=
\set{r[\var(R_i)]:r\in E}$, where
for  $X\subset
\set{x_1,\ldots,x_d}$, $r[X]:X\to \mathbb R$ denotes 
the restriction of $r$ to~$X$.
 Clearly, $|R_i(\str D)|=|\pi_i(E)|$.
Let $Q$ be the natural join query 
consisting of all the $R_i$'s. Then $E\subset Q(\str D)$,
and $|Q(\str D)|\le\prod_{j=1}^d|\pi_j(E)|^{w_i}$ by
Corollary~\ref{cor:agm-full}, where $(w_i)_{i=1}^d$
is any fractional edge covering of the hypergraph 
with vertices $x_1,\ldots,x_d$ and hyperedges
which are complements of singletons. 
In particular, since every vertex belongs to exactly
$d-1$ hyperedges,
taking $w_i=1/(d-1)$ yields a fractional
edge covering, proving  the discrete variant of Theorem~\ref{thm:lw}.
The continuous variant can 
be derived by approximation, in a standard way.

Theorem~\ref{thm:simple-stat}, in principle,
can be used to formulate a stronger inequality than the discrete Loomis-Whitney inequality (or other geometric
inequalities listed above), for sets $E\subset\mathbb R^n$
which satisfy given functional dependencies:
we say that $E$ satisfies a functional dependency 
$I\mapsto j$ (where $I\subset\set{1,\ldots,n},j\in\set{1,\ldots,n}$) if for every $v,w\in E$ such that $v_i=w_i$ for all $i\in I$,
it is the case that $v_j=w_j$.
\begin{corollary}
  Let $E\subset \mathbb R^d$ be a finite set, satisfying
  a given set of functional dependencies $\Ff$.
  Then $|E|\le \beta,$
   where $\beta$ is
  the maximal value of $v_X$,
   for $X=\set{1,\ldots,d}$ and $v$ ranging over $\overline{\Gamma_X^*}$ and satisfying:
   $$\begin{cases}
     v_{X-\set{i}}\le  \log|\pi_j(E)|&\text{ for }j\in X,\\
     v_{I\cup \set{j}}=v_I &\text{ for every fd $I\mapsto j$ in $\cal F$.}
   \end{cases}$$
\end{corollary}
Although the problem of computing the value $\beta$ is
not addressed in this paper, Theorem~\ref{thm:groups}
(or rather, its more precise variant, corresponding to Theorem~\ref{thm:simple-stat}) provides a description of worst-case sets $E$.

\paragraph{Other results}
A generalization of Theorem~\ref{thm:optimization} to conjunctive queries (with projections) is possible, and we discuss such a result in Section~\ref{sec:general}.
Also, we describe a crude algorithm for evaluating $Q(\str D)$ in Section~\ref{sec:complexity}.

 \paragraph{Outline of the paper}
After introducing notation, definitions, and preliminary observations in Section~\ref{sec:prelim}, we recall the entropy (upper) bound
for $\alpha(Q)$ from~\cite{gottlob} in Section~\ref{sec:upper}. Then, in Section~\ref{sec:lower} we
 present several lower bounds for $\alpha(Q)$: we recall the coloring bound from~\cite{gottlob}, and later improve it to vector space colorings, and group systems. Finally, in Section~\ref{sec:tight},
 we show that the group system bound matches the entropy bound, using a construction from~\cite{Lun02}.
 This proves Theorem~\ref{thm:optimization} and Theorem~\ref{thm:groups}.
 In Section~\ref{sec:general} we show how to generalize Theorem~\ref{thm:optimization} to queries with projections,
 In Section~\ref{sec:complexity}, we describe some basic results concerning the worst-case complexity of computing $Q(\str D)$, for a fixed query $Q$ and given database~$\str D$.
 
 \paragraph{Acknowledgements} 
 We are grateful to Dan Suciu for introducing
 the problem to us, and to Adam Witkowski for fruitful discussions.

\section{Preliminaries}\label{sec:prelim}
To fix notation, we recall some notions concerning databases,
and entropy.
\paragraph{Notation}
We assume a fixed \emph{schema} $\Sigma$,
which specifies a finite set of \emph{attributes} $\var(\Sigma)$, a finite set of \emph{relation names}, and for each relation name $R$, a finite set $\var(R)\subset\var(\Sigma)$ of attributes of $R$. 
If $X$ is a set of attributes, then a \emph{row} with attributes $X$ is a function $r$ assigning to each $x\in X$ some value $r[x]$. If $r$ is a row with attributes $X$ and $Y\subset X$ then by $r[Y]$ we denote the restriction of $r$ to $Y$.
A \emph{table} with attributes $X$ is a finite set of rows with attributes $X$.
A \emph{database} $\str D$ over $\Sigma$ specifies for each relation name $R$ in $\Sigma$ a table with attributes $\var(R)$. 
A \emph{natural join query} is a set $Q$ of relation names in $\Sigma$; we denote $\var(Q)=\bigcup_{R\in Q}\var(R).$
Such a query  can be applied to a database $\str D$, yielding as result the table $Q(\str D)$ consisting of those rows $r$ with attributes $\var(Q)$ such that $r[{\var(R)}]\in R(\str D)$ for every $R\in Q$.

We say that a database $\str D$ \emph{satisfies} a \emph{functional dependency} $R:X\mapsto x$ -- where $X$ is a set of attributes  and $x$ is a single attribute -- if for any two rows $u,v$ of $R(\str D)$, $u[X]=v[X]$ implies $u[x]=v[x]$.

For the rest of this paper, fix a schema $\Sigma$  and a set of functional dependencies $\Ff$. Every database $\str D$ is assumed to be over this schema, and to satisfy  $\Ff$.
Define $\alpha(Q)$ as the smallest value~$\alpha$ for which there exists a constant $c$  such that~\eqref{eq:main} holds for all databases $\str D$ over $\Sigma$ which satisfy the functional dependencies in $\Ff$. For convenience, we define $|\str D|$ to be the maximal size of a relation in $\str D$. 
Since we allow a multiplicative constant in~\eqref{eq:main},
 defining $|\str D|$ as the sum of the sizes of the relations in $\str D$ would give an equivalent definition of $\alpha(Q)$.

\begin{remark}
Observe that since $Q$ is a natural join query and in the definition of $\alpha(Q)$ we are interested in maximizing $|Q(\str D)|$ while keeping $|\str D|$ bounded, we may assume that if $R:X\mapsto x$ is a functional dependency in $\Ff$, then also  $S:X\mapsto x$ is  a functional dependency in $\Ff$, for every relation name $S$ such that $X\subset \var(S)$. Therefore, we may simply write that $\Ff$ contains the functional dependency $X\mapsto x$
instead of writing $R:X\mapsto x$.
\end{remark}

For a database $\str D$ (over $\Sigma$, satisfying $\Ff$),  denote
\begin{align}\label{eq:alpha}\tag{$\alpha$}
\alpha(Q,\str D)=\frac{\log|Q(\str D)|}{\log|\str D|}.
\end{align}

\paragraph{Convention}
Throughout this paper we will define several real-valued parameters of the form $\gamma(Q,x)$, where $Q$ is a query and $x$ is some object. For a fixed query $Q$, we denote by $\sup_x \gamma(Q,x)$ the supremum, and by $\limsup_x \gamma(Q,x)$ the limit superior over all values $x$, for which the value $\gamma(Q,x)$ is defined. In particular, $\limsup_x\gamma(Q,x)$ is the smallest value in $\R\cup\set{-\infty,+\infty}$ such that for every  real $\eps>0$, there are only finitely many $x$'s such that  $\gamma(Q,x)$ is defined and larger than ${\gamma(Q)+\eps}$. 

\paragraph{Limit superior vs. supremum} 
The following simple lemma will simplify 
several formulations and proofs throughout this paper.
\begin{lemma}\label{lem:sup}
  For a natural join query $Q$, $$\alpha(Q)=\limsup_{\str D}{\alpha(Q,\str D)}=\sup_{\str D}\alpha(Q,\str D).$$
\end{lemma}
\begin{proof}Let $d=\limsup_{\str D}{\alpha(Q,\str D)}.$ First we show that $d=\alpha(Q)$.
  By definition, for every $\eps>0$ there are finitely many databases $\str D$ for which
  $$\frac{\log|Q(\str D)|}{\log|\str D|}>  d+\eps,$$
  so $|Q(\str D)|\le |\str D|^{d+\eps}$
  for almost all $\str D$. By choosing a large enough constant $c$, we have that 
  $$|Q(\str D)|\le c\cdot |\str D|^{d+\eps}$$ for all databases $\str D$. Hence, $\alpha(Q)\le d+\eps$, for every $\eps>0$, proving $\alpha(Q)\le d$.
    The inequality $d\le \alpha(Q)$ is proved similarly.

  To show that $\sup_{\str D}\alpha(Q,\str D)\le\limsup_{\str D}\alpha(Q,\str D)$, we use the following construction. For a database $\str D$ and a natural number $n$, let $\str D^n$ be the database defined so that the rows of $R(\str D^n)$ are $n$-tuples of rows of $R(\str D)$, and for such a row $r=(r_1,\ldots,r_n)$, we define $r[x]=(r_1[x],\ldots,r_n[x])$ for an attribute $x\in\var(R)$.
It is easy to check that $\str D^n$ satisfies the same functional dependencies as $\str D$,
  and that $|\str D^n|=|\str D|^n$ and  $|Q(\str D)|=|Q(\str D)|^n$. In particular, $\alpha(Q,\str 
D^n)=\alpha(Q,\str D)$. 
It follows that if $|\str D|>1$, then by choosing $n$ arbitrarily large, 
we have arbitrarily large databases $\str D^n$ with $\alpha(Q,\str D^n)=\alpha(Q,\str D)$.
 Therefore, $\limsup_{\str D}\alpha(Q,\str D)\ge \sup_{\str D}\alpha(Q,\str D)$, the other inequality being obvious.
\end{proof}

\paragraph{Entropy}
In this paper, we only consider random variables taking finitely many values. Formally, a random variable $X$ is a measurable function $X:\Omega\to V$ from a fixed probability space $(\Omega,\mathbb P)$ of events to a finite set $V$. In this paper, however, it is 
not harmful to assume that $\Omega$ is a finite probability space, in which 
case every function $X:\Omega\to V$ is a random variable. By $\im X\subset V$ we denote the set of 
values $v$ such that $\Pr{X=v}>0$, where $\Pr{X=v}$
is a shorthand for $\Pr{\set{\omega\in\Omega:X(\omega)=v}}$.

For a random variable $X$ taking values in a finite set $V$, define the \emph{entropy} of $X$ as
$$H(X)=-\sum_{v\in \im X}p_v\log  p_v,$$
where $p_v=\Pr{X=v}$. Clearly, the entropy of $X$ only depends on the distribution of $X$. Also, the maximal possible entropy of a random variable with values in a finite set $V$ is
 equal to $\log|V|$, and is attained by the uniform distribution on $V$, as follows from Jensen's inequality applied to the convex function $-\log(x)$.

 According to Shannon's source coding theorem, the value $H(X)$ has the following  characterization (up to an additive error of $1$): how many questions on average does one need to ask to determine the  value of a random variable $X$, where each question is a question of the form ``does $X$ belong to $U$?'' (where $U\subset \im X$)? Here we mean the minimum, over all strategies against a given distribution of $X$, of the average  number of questions.

\section{Upper bound}
\label{sec:upper}

We start with presenting an upper bound on $\alpha(Q)$, 
which we call the \emph{entropy bound}.
This bound 
is essentially from~\cite{gottlob}.

\paragraph{The entropy bound}
Fix a natural join query $Q$.
Let $U$ be a random variable $U$ taking as values rows with attributes $\var(\Sigma)$.
For a set of attributes $X\subset \var(\Sigma)$, define $U[X]$ to be 
the random variable whose value is the restriction of the value of $U$ to the set of attributes $X$. 
In particular, $U[X]$ is a random variable whose values are rows with attributes $X$, and $\im{U[\var(R)]}$ is a table with attributes $\var(R)$.
We say that $U$ satisfies a functional dependency $Y\mapsto x$ if the table $\im U$ satisfies the functional dependency $Y\mapsto x$.

\begin{lemma}\label{lem:functional-entropy}
The random variable $U$ satisfies a functional dependency $Y\mapsto x$ if and only if 
\begin{align}\label{eq:ent-fd}
\ent(U)_Y=\ent(U)_{Y\cup\set x}.  
\end{align}

\end{lemma}
  We remark that in information theory, \eqref{eq:ent-fd} can be expressed using conditional entropy as ${H(\ U[x]\ |\ U[Y]\ )=0}$.

\begin{proof} By unraveling of definitions. Obviously~\eqref{eq:ent-fd} holds if and only if $U[Y]$ and $U[Y\cup\set x]$ have the same distributions. This however means that there exist no elements $r, c \neq  d$ such that $\Pr{U[Y]=r, U[x]=c}>0$ and  $\Pr{U[Y]=r, U[x]= d}>0$. This is the definition of functional dependency $Y\mapsto x$ for random variables. 
\end{proof}

 For a random variable $U$ which satisfies every functional dependency in $\Ff$ we define
\begin{align}\label{eq:H}\tag{H}
H(Q,U)&=\frac{H(U[\var(Q)])}{\max_{R\in\Sigma}H(U[\var(R)])},
\end{align}
and let $H(Q)=\sup_U H(Q, U)$.

Observe that a random variable $U$
taking as values rows with attributes $X=\var(Q)$
is the same thing as a tuple $(U_x)_{x\in X}$ of random variables. With this observation, Lemma~\ref{lem:functional-entropy} allows us to 
  characterize $H(Q)$ in terms of the entropy cone $\overline{\Gamma_X^*}$.
  
\begin{prop}\label{pro:optimization}
  The value $H(Q)$ is equal to the value described by the optimization problem from Theorem~\ref{thm:optimization}.  
\end{prop}
\begin{proof}By definition of the entropy cone $\overline{\Gamma_X^\ast}$.  
\end{proof}

  Therefore, to prove Theorem~\ref{thm:optimization},
  it remains to show that $\alpha(Q)=H(Q)$. 
  Lemma~\ref{lem:H-alpha} below 
 shows one of the two inequalities, 
   by employing an observation from~\cite{gottlob},
  
\begin{lemma}\label{lem:H-alpha}Let $Q$ be a natural join query. Then
  $${H(Q)\ge \alpha(Q).}$$
  \end{lemma}
  \begin{proof}\label{pf:}
    For a database $\str D$, define 
    a random variable denoted $U_\str D$ which chooses uniformly at random a row of $Q(\str D)$.
     By definition of a natural join query, the values of $U_\str D[R]$ are rows of $R(\str D)$. 
    Then $H(Q,U_\str D)\ge \alpha(Q,\str D)$ for every database $\str D$, since
$H(U_\str D)=\log|Q(\str D)|$ and
$H(U_\str D[R])\le \log|R(\str D)|$ by the fact that the uniform distribution maximizes entropy. This, together with Lemma~\ref{lem:sup}, proves Lemma~\ref{lem:H-alpha}.
  \end{proof}
  
In Sections~\ref{sec:lower} and~\ref{sec:tight} we prove the remaining inequality $H(Q)\le \alpha(Q)$, thus finishing the proof of Theorem~\ref{thm:optimization}.

\section{Lower bounds}
\label{sec:lower}
The paper~\cite{gottlob} also provides a lower bound for $\alpha(Q)$, using colorings, which we recall below for completeness.
We then improve this bound to vector space colorings, and finally, to group systems. In Section~\ref{sec:lower},
fix a natural join query $Q$ over a schema $\Sigma$, and a set of functional dependencies $\Ff$.

\subsection{Colorings}
A \emph{coloring} of $Q$ is a function
 $f$ assigning finite sets to $\var(Q)$.
 We say that $f$ satisfies a functional dependency $X\mapsto x$ if $f(x)\subset f(X)$, where $f(X)$
 denotes $\bigcup_{y\in X}f(y)$.
For a coloring $f$
 of $Q$ which satisfies all functional dependencies in $\Ff$, define
\begin{align}\label{eq:C}\tag{C}
C(Q,f)=\frac{|
f(\var(Q))|}{\max_{R\in\Sigma} |f(\var(R))|},
\end{align}
and let $C(Q)=\sup_f C(Q,f)$.
 
\begin{lemma}[\cite{gottlob}]\label{lem:C-alpha}
  Let $Q$ be a natural join query. Then
  $\alpha(Q)\ge C(Q).$
\end{lemma}

\begin{proof}\label{pf:}
Consider a coloring $f$ of $Q$ satisfying the functional dependencies $\Ff$.
We construct a database $\str D$ with $\alpha(Q,\str D)\ge C(Q,f)$.
Let $C=f(\var(Q))$ be the set of all colors used by $f$.

Choose a set $N$ with $|N|>1$, and consider the table $T=N^C$ with attributes $C$.
Define the database $\str D$ so that for each relation name $R$,  $$R(\str D)=\set{r[f(\var(R))]: r\in T}.$$
 Then it is not difficult to check that:
\begin{itemize}
	\item The database $\str D$ satisfies the required functional dependencies,
	\item For every relation name $R$, $|R(\str D)|=|N|^k$, where $k=|f(\var(R))|$,
	\item $|Q(\str D)|=|N|^{|C|}$.
\end{itemize}
This yields that $\alpha(Q,\str D)= C(Q,f)$,
and hence $\alpha(Q)\ge C(Q)$ by Lemma~\ref{lem:sup}.
\end{proof}

It is shown in~\cite{gottlob} that the value $C(Q)$ can be computed by a linear program. In the case without  functional dependencies, this program is dual to the program for $\agm(Q)$, so $C(Q)=\agm(Q)=\alpha(Q)$. 

However, in the presence of functional dependencies, there are queries $Q$ for which $\alpha(Q)> C(Q)$, as shown in the paper~\cite{gottlob}, by elaborating an example proposed by D{\'a}niel Marx. Interestingly, this construction uses Shamir's secret sharing scheme, which is based on the fact a polynomial of degree $k$ is uniquely determined by any of its $k$ values.
There is another secret sharing scheme, Blakley's scheme, which employs vector spaces rather than polynomials, and the fact that any point in a $k$-dimensional vector space is uniquely determined by a $k$-tuple of hyperspaces in general position which contain it. More generally, we have the following lemma. 

If $V$ is a vector space, $W$ is its subspace and $v\in V$, then $W+v=\set{w+v:w\in W}$
is the unique hyperspace in $V$ that is parallel to $W$
and contains $v$.

\begin{lemma}\label{lem:secret}Fix a vector space $V$,
a family $(V_x)_{x\in X}$ of subspaces of $V$,
and a subspace  $V_0\subset V$ such that $V_0\supseteq \bigcap_{x\in X}V_x$. 
For any given $v\in V$, the hyperspaces $(V_x+v)_{x\in X}$ determine $V_0+v$, i.e., for every two vectors $v,w\in V$,
if $V_x+v=V_x+w$ for all $x\in X$, then $V_0+v=V_0+w$.
\end{lemma}
\begin{proof}If $V_1\subset V_0$ then $V_1+v\subset V_0+v$,
    and $V_0+v$ is determined by $V_1+v$ as follows:
    $V_0+v=\set{w+w':w\in V_0,w'\in V_1}$. In words, $V_0+v$
    is the unique hyperspace parallel to $V_0$
    that contains $V_1+v$.

    Applying this observation to $V_1=\bigcap_{x\in X}V_x$
    yields that $V_0+v$ is determined by $(\bigcap_{x\in X}V_x)+v=\bigcap_{x\in X} (V_x+v)$.    
\end{proof}

The above lemma leads to a ``multi-secret'' sharing scheme,
in which every participant has a publicly known subspace $V_x$ of $V$, and his secret hyperplane $V_x+v$, where $v\in V$ is fixed and unknown.
A set of participants $Y$ can gather 
and determine the secret of the participant $x$ whenever $V_x\supseteq \bigcap_{y\in Y}V_y$.
This secret sharing scheme leads us to construction providing a tighter upper bound, which we describe below.

\subsection{Vector space colorings}
We consider vector spaces over a fixed finite field $\k$. If $V$ is a vector space, $X$ is a set, and $V_x$ is a subspace of  $V$ for $x\in X$, then by $\sum_{x\in X} V_x$ we denote the smallest subspace of $V$ containing every $V_x$, for $x\in X$.
For a subspace $W$ of a vector space $V$, by $\codim_V W$
we denote $\dim V-\dim W=\dim (V/W)$,
where $V/W$ denotes the quotient space, $V/W=\set{W+v:v\in V}$.

A \emph{vector space coloring} of $Q$
is a pair $\Vv=(V,(V_x)_{x\in \var(Q)})$, where $V$ is a vector space and $(V_x)_{x\in\var(Q)}$ is a family of its subspaces. For such a coloring, define 
$V_Y=\sum_{x\in Y}V_x$ for $Y\subset \var(Q)$.
We say that $\Vv$ \emph{satisfies} a functional dependency
$Y\mapsto x$ if $V_x\subset V_Y$.
For a vector space coloring $\Vv$ satisfying all the functional dependencies in $\Ff$, define
\begin{align}\label{eq:vc}\tag{VC}
\vc(Q,\Vv)=\frac{\dim (V_{\var(Q)})}{\max_{R\in\Sigma} \dim(V_{\var(R)})},	
\end{align}
Let	$\vc(Q)=\sup_\Vv \vc(Q,\Vv)$.
 
\begin{prop}\label{prop:vc}Let $Q$ be a natural join query. Then	
	\begin{align}
	\alpha(Q)\ge \vc(Q)\ge C(Q).
	\end{align}
\end{prop}
The inequality $\vc(Q)\ge C(Q)$ is obtained by 
defining for a coloring $f$ a vector space coloring
 $\Vv$ with $V=\k^C$, $V_x=\k^{f(x)}\subset \k^C$,
 where $C=\bigcup_{x\in \var(Q)} f(x)$,
 and $\k^{f(x)}$ embeds into $\k^C$ in the natural way,
 by extending a vector with zeros on coordinates in $C-f(x)$. 
  It is easy to see that
  $\Vv$ satisfies the same functional dependencies as $f$,
and that $\vc(Q,\Vv)=C(Q,f)$.  This proves $\vc(Q)\ge C(Q)$.
To prove the bound $\alpha(Q)\ge \vc(Q)$, we pass to dual vector spaces, as described below.
 
A \emph{vector space system} over $Q$
is a pair $\Vv=(V,(V_x)_{x\in \var(Q)})$, where $V$ 
is a vector space and $(V_x)_{x\in \var(Q)}$ is a family of its subspaces. For such a system, define $V_Y=\bigcap_{y\in X}V_y$ for $Y\subset \var(Q)$, and
 say that $\Vv$ satisfies a functional dependency $Y\mapsto x$ if $V_x\supseteq V_Y$.

For a vector space system $\Vv$ which satisfies all the functional dependencies in $\Ff$, define 
	\begin{align}\label{eq:vc*}\tag{VC$^*$}
\vcs(Q,\Vv)=\frac{\codim_V
(V_{\var(Q)})
}{\max_{R\in\Sigma} \codim_V(V_{\var(R)})}.
\end{align}	
Finally, let $\vcs(Q)=\sup_\Vv \vcs(Q,\Vv)$.

\begin{lemma}\label{lem:bij}
	There is a bijection between vector space colorings and vector spaces systems,
which  maps a vector space coloring $\Vv$ to a vector space system $\Vv^*$ such that $\vcs(Q,\Vv^*)=\vc(Q,\Vv)$.
In particular,	$\vc(Q)=\vcs(Q).$
\end{lemma}

\begin{proof}\label{pf:}	
If $V$ is a vector space, let $V^*$ denote its algebraic dual, i.e., the space of all linear functionals from $V$ to $\k$. If $L$ is a subspace of $V$, then let $L^\bot\subset V^*$ denote the set of functionals $f\in V^*$ which vanish on $L$. Then 
\begin{align}
(L_1\cap L_2)^\bot&=L_1^\bot+L_2^\bot\label{eq:dualsum}\\
L_1\subset L_2&\iff L_1^\bot\supseteq L_2^\bot\label{eq:dualsubset}\\
 \dim(L^\bot)&=\codim_V(L).\label{eq:dualdim}
\end{align}

For a vector space coloring $\Vv=(V,(V_x)_{x\in \var(Q)})$  let $\Vv^*=
(V^*, (V_x^\bot)_{x\in \var (Q)})$. Using the facts~\eqref{eq:dualsum},\eqref{eq:dualsubset},\eqref{eq:dualdim}, it is easy to check that the mapping $\Vv\mapsto \Vv^*$ yields a bijection with the required properties.
\end{proof}
\begin{proof}[Proof of Proposition~\ref{prop:vc}]
	
	We prove the inequality $\alpha(Q)\ge \vcs(Q)$.
	Let $\Vv=(V,(V_x)_{x\in \var(Q)})$ be a vector space system. 
    	We construct a  database $\str D$ satisfying
  \begin{align}\label{eq:alpha-vc}
     \alpha(Q,\str D)= \vcs(Q,\Vv).
  \end{align}

	For a relation name $R$ and 
	a vector $v\in V$,
	let $r_v$ be the row such that $r_v[x]=V_x+v\in V/V_x$
	for every attribute $x\in\var(R)$.
	Define $\str D$ by setting
	$$R(\str D)=\set{r_v:		v\in V},$$
	for every relation name $R$. It is easy to verify that:
\begin{itemize}
	\item The database $\str D$ satisfies the functional dependencies. This follows from~Lemma~\ref{lem:secret}.
	\item For each relation name $R$,
  consider the mapping $f_R:V\to R(\str D)$, where $f_R(v)=r_v$. Clearly, the mapping is onto $R(\str D)$. To compute the size of its image, we analyse the kernel of $f_R$ and observe that $\set{w:r_w=r_v}=V_{\var(R)}+v$
  for every $v\in V$.
  In particular, $|R(\str D)|  =|V|/|V_{\var(R)}|=|\k|^{\codim_V V_{\var(R)}}$.
  
    \item Similarly, we verify that    
    $|Q(\str D)|=|V|/|V_{\var(Q)}|=|\k|^{\codim_V V_{\var(Q)}}$.
\end{itemize}	
Equation~\eqref{eq:alpha-vc} follows.
\end{proof}

\subsection{Group systems}\label{sec:group_coloring}
We relax the notion of a vector space system, by considering finite groups, as follows.
Let $G$ be a finite group and $(G_x)_{x\in \var(Q)}$ be a family of its subgroups. 
For a set of attributes $X\subset \var(Q)$, denote by $G_X$ the group $G_X=\bigcap_{x\in X}G_x$, and by $G/G_X$ the
space of (left) cosets, $\set{g\cdot G_X:g\in G}$.
We call the pair ${\Gg}=(G,(G_x)_{x\in X})$ 
a \emph{group system} for $Q$, and say that it satisfies a functional dependency $X\mapsto x$
if $G_X\subseteq G_x$.
For  a group system $\Gg$ satisfying all the functional dependencies in $\cal F$, define
	\begin{align}\label{eq:g}\tag{GC}
\gc(Q,{\Gg})=\frac{\log
|G/G_{\var(Q)}|}{\max_{R\in\Sigma}(\log|G/G_{\var(R)}|)},
\end{align}	
and let $\gc(Q)=\sup_{\Gg}\gc(Q,\Gg)$.
 
 \begin{prop}\label{prop:lower} Let $Q$ be a natural join query. Then
	 \begin{align}
 	\alpha(Q)\ge \gc(Q)\ge \vcs(Q)=\vc(Q)\ge C(Q).
	 \end{align}
 \end{prop}
 \begin{proof}
   Clearly, $\gc(Q)\ge \vcs(Q)$, since every vector space system is a group system.
   
   The proof of the inequality $\alpha(Q)\ge \gc(Q)$ is analogous to the proof of Proposition~\ref{prop:vc}. From a group
   system $\Gg$, we construct a database $\str D$.
   Here, we use cosets $g\cdot G_x$
   instead of translates $V_x+v$, and the set of cosets $G/G_x$
   instead of quotients $V/V_x$.
   By following the same construction of $\str D$, we get the following:
   \begin{itemize}
   	\item The database $\str D$ satisfies the functional dependencies. Here we use a lemma analogous to Lemma~\ref{lem:secret}, which holds for groups as well.
   	\item For each relation name $R$ and vector $v$,
     $\set{h:h_r=g_r}=G_{\var(R)}\cdot g$.
     In particular, $|R(\str D)|=|G|/|G_{\var(R)}|$;
  
       \item $|Q(\str D)|=|G|/|G_{\var(Q)}|$.
   \end{itemize}
This proves $\alpha(Q,\str D)\ge \gc(Q,\Gg)$, and hence
$\alpha(Q)\ge \gc(Q)$ by Lemma~\ref{lem:sup}.
 \end{proof}

\begin{remark}\label{rem:symmetric}
  The database $\str D$ constructed in the above proof is $G$-symmetric. Indeed, the values appearing in the database are  cosets the form $g\cdot G_x$ (where $x\in\var(Q)$)  
  and $G$ acts (from the left) on such cosets in the obvious way. Moreover, the action of $G$ on each table $R(\str D)$ is isomorphic to the action of $G$
  on $G/G_{\var(R)}$, in particular, it is transitive.  
  
  Also, if $\str D$ is $G$-symmetric and $\str D^n$ is defined as in the proof of Lemma~\ref{lem:sup}, then $\str D^n$ is $G^n$-symmetric.
\end{remark}

\section{Tightness}\label{sec:tight}
In Sections~\ref{sec:upper} and~\ref{sec:lower}
we have shown that
for every natural join query $Q$,
 $$H(Q)\ge \alpha(Q)\ge \gc(Q).$$
In this section, we close the circle by proving the following.
\begin{prop}\label{prop:upper}
  Let $Q$ be a natural join query. Then
  	 \begin{align}
   	\gc(Q)\ge H(Q).
  	 \end{align}
\end{prop}
\noindent Together with Proposition~\ref{prop:lower} and Lemma~\ref{lem:H-alpha}, this proves that 
$H(Q)=\alpha(Q)=\gc(Q)$. As noted in Proposition~\ref{pro:optimization}, this gives Theorem~\ref{thm:optimization}. Moreover, from the observation in Remark~\ref{rem:symmetric}, it follows
that the bound $\alpha(Q)$ can be approximated by (arbitrarily large) symmetric databases, proving Theorem~\ref{thm:groups}.

All the necessary ideas to prove the proposition are present in the proof of main theorem from \cite{ChanY02}  in the version presented in the paper \cite{Lun02}. However, we cannot directly apply that theorem, since we need to keep track of the functional dependencies.
In the rest of Section~\ref{sec:tight}, we present a self-contained proof 
of Proposition~\ref{prop:upper}, following the ideas of~\cite{Lun02}.
For the rest of this section, fix a random variable $U$, taking values in a finite set of rows with attributes $X$,
and satisfying the functional dependencies $\cal F$.

We say that a random variable $V$ is \emph{rational} if for every value $v\in\im V$, the probability that $V$ achieves $v$ is a rational number.   
%
%
%


\begin{lemma}\label{lem:przyblizanie}
   For every number $\eps >0$ there exists a rational random variable $V$ 
   satisfying the same functional dependencies as $U$, and such that $\|\ent(V)-\ent(U)\|<\eps$ with respect to the euclidean norm
   on ${\mathbb R}^{P(X)}$.
\end{lemma}
\begin{proof}
  Let $T=\im U $. Observe that every random variable $V$ with values in $T$
  satisfies the same functional dependencies as $U$.
  Denote by ${\cal D}(T)$  the set of probability
  distributions on $T$.
  Consider the mapping $\ent:{\cal D}(T)\to {\mathbb R}^{P(X)}$,
  which maps a probability distribution $D\in {\cal D}(T)$ to the entropy vector
  $\ent(V)\in {\mathbb R}^{P(X)}$  of any random variable $V$ with distribution $D$.
As is clearly visible from the explicit formula for this mapping, it is continuous. We conclude the lemma by observing that among the set of all probability  distributions on $T$, distributions with rational values form a dense subset.
%
\end{proof}

To prove Proposition~\ref{prop:upper}, we proceed as follows. 
For each rational random variable $U$ satisfying the given functional dependencies,
we will find a sequence of 
group systems $\Gg_k$ satisfying the functional dependencies $\cal F$, and such that
  \begin{align}
  \label{eq:spade}\lim_{k\rightarrow \infty}\gc(Q,\Gg_k) = H (Q, U).
  \end{align}
From that, Proposition~\ref{prop:upper} follows:
\begin{multline*}
H(Q)=\sup_{U}{H(Q, U)} \stackrel{\textit{Lem.\ref{lem:przyblizanie}}}{=}\sup_{U\text{ rational}}{H(Q, U)} =\\
  \sup_{U\text{ rational}}{H(Q, U)}
 \stackrel{\textrm{\eqref{eq:spade}}}{\leq} \sup_\Gg{\gc(Q,\Gg)=\gc(Q)}.
\end{multline*}

From now on, let $U$ be a rational random variable
satisfying the functional dependencies~$\Ff$.
 We will show that there exists a sequence of group systems witnessing~\eqref{eq:spade}.

Let $q\in \mathbb{N}$ be a natural number such that for each row $r\in \im{U}$ the probability $\Pr{U=r}$ can be represented as a rational number with denominator $q$. For $k=q,2q,3q,\ldots$ let $A_k$ be a matrix whose columns are indexed by attributes,
containing exactly $k\cdot \Pr{U=r}$ copies
of the row $r$, for every row $r\in\im U$. Notice that  $k\cdot \Pr{U=r}$ is always a natural number, and that $A_k$ has exactly $k$ rows in total.

 Let $G^k$ denote the group of all permutations
of the rows of $A_k$.
For a set of attributes $Y\subset X$, let $G^k_Y$
denote the subgroup of $G^k$ which 
stabilizes the submatrix $A_k[Y]$ of $A_k$, i.e.,
  $$G^k_Y=\{\sigma\in G^k_Y ~|~ \sigma(r)[y]=r[y]\text{ for each $r\in A_k$ and $y\in Y$}\}.$$
Denote $G^k_{\set x}$ by $G^k_{ x}$. In particular, $G^k_Y=\bigcap_{y\in Y}G^k_y$.
The following lemma is  immediate.

\begin{lemma}\label{lem:satfundep}
 Suppose that $U$ satisfies the functional dependency $Y\mapsto x$. Then $G_Y^k\subseteq G_x^k$. 
\end{lemma}


We define $\Gg_k$ as $(G^k,(G^k_x)_{x\in X})$. By Lemma \ref{lem:satfundep}, $\Gg_k$ is a  group system
satisfying the functional dependencies~$\cal F$.
It remains to prove~\eqref{eq:spade}.

Fix a set of attributes $Y\subseteq X$.
For a row $r\in \im{U[Y]}$, let  $p_r$ denote $\Pr{U[Y]=r}$.
Since $r$ occurs exactly $k\cdot p_r$ times as a row of $A_k[Y]$, it follows that
$$ |G^k_Y|=\prod_{r\in \im{U[Y]}} (k\cdot p_r)!.$$

Using Stirling's approximation we get that
\begin{multline*}
 \lim_{k \rightarrow \infty}\frac{1}{k} \log\left(\frac{|G^k|}{|G^k_Y|}\right)=
\lim_{k \rightarrow \infty}\frac{1}{k} \log\left(\frac{k!}{\prod_{r\in \im{U[Y]}} (k\cdot p_r)!}\right)=\\
\lim_{k \rightarrow \infty}\frac{1}{k}\left(  k\log(k)-{\sum_{r\in \im{U[Y]}} (k\cdot p_r)\log \left(k\cdot p_r\right)}\right)=\\
\lim_{k \rightarrow \infty}\frac{1}{k}\left(  -{\sum_{r\in \im{U[Y]}} (k\cdot p_r)(\log \left(k\cdot p_r-\log{k})\right)}\right)=\\
\lim_{k \rightarrow \infty}\frac{1}{k}  \cdot(-k)\cdot{\sum_{r\in \im{U[Y]}}  p_r\log p_r}=
\ent(U)_Y.
\end{multline*}

In particular, $$\gc(Q,\Gg^k)=
\frac{\frac 1k\log (|G^k|/|G^k_X|)}{\max_{R\in Q}\frac 1k\log (|G^k|/|G^k_{\var R}|)}\stackrel{k\rightarrow\infty}{\longrightarrow}	$$
$$\stackrel{k\rightarrow\infty}{\longrightarrow}
\frac{H(U)}{\max_{R\in Q}H(U[\var(R)])}=H(Q,U).
$$
This yields~\eqref{eq:spade} proving Proposition~\ref{prop:upper},
which together with Lemma~\ref{lem:H-alpha} gives $H(Q)=\gc(Q)=\alpha(Q)$.
By Proposition~\ref{pro:optimization}, this finishes the proof
of Theorem~\ref{thm:optimization}.
The more general Theorem~\ref{thm:simple-stat} is proved similarly. Moreover, Theorem~\ref{thm:groups}
follows from Remark~\ref{rem:symmetric}.
\section{General conjunctive queries}\label{sec:general}

The previous sections concern natural join queries: conjunctive queries without existential quantifiers (or projections), in which the variables name
 coincides with the name of the attribute of the relation
 in which it appears
  (in particular, the same variable name cannot occur in the scope of one conjunct,
and there are no equalities).
In this section, we discuss how to treat arbitrary conjunctive queries. 

\subsection{Set semantics}
Define a \emph{natural conjunctive query}  to be a query of the form $Q=\exists Y~ Q'$, where $Q'$ is a natural join query over the schema $\Sigma$, and $Y\subset \var(\Sigma)$. The set of \emph{free variables} of $Q$
is $\var(Q)=\var(\Sigma)-Y$.
 For a database $\str D$ over the schema $\Sigma$, define $Q(\str D)$ as $T[\var(Q)]$, where $T=Q'(\str D)$. In other words, $Q(\str D)$ is the table $Q'(\str D)$ restricted to the free variables of $Q$. This is the so-called \emph{set-semantics}, since the result $T[\var(Q)]$
 is a set of rows, i.e., each row occurs either 0 or 1 times. The alternative \emph{bag-semantics} is discussed in Section~\ref{sec:bag}.

  As explained in \cite{gottlob} for each conjunctive query $Q$ there exists a  natural conjunctive join query  $S$, such that $\alpha(Q)=\alpha(S)$. Such $S$ can be constructed in a  purely syntactical way from $Q$ by the chase procedure. Because of this, we only consider natural conjunctive queries.
 
 The definition of $\alpha(Q)$
can be lifted without modification to natural conjunctive queries.
The generalization of Theorem \ref{thm:optimization} has the expected form:
\begin{theorem}  \label{thm:arbitrary_query}
  Fix a relational schema $\Sigma$ and a set of functional dependencies $\cal F$.  Let $Q$ be a natural conjunctive  query  over the schema $\Sigma$.
    Then  $\alpha(Q)$ is equal to the maximal value of $v_{\var(Q)}$,
    for $v$ ranging over $\overline{\Gamma_{\var(\Sigma)}^*}$ and satisfying:
    $$\begin{cases}
      v_{\var(R)}\le 1&\text{ for }R\in Q,\\
      v_{Z\cup \set{z}}=v_Z &\text{ for every fd $Z\mapsto z$ in $\cal F$.}
    \end{cases}$$
\end{theorem}

The proof of the lower bound is exactly the same as the proof of Theorem \ref{thm:optimization}. However the proof of the upper bound has to be modified slightly, as described below. 

For a query $Q=\exists Y.\ Q'$, define $H(Q,U)$ and $H(Q)$ as in Section~\ref{sec:upper}, where $R$ in the maximum ranges over $Q'$ rather than $Q$.
We then have the following analogue of Lemma~\ref{lem:H-alpha}, proving the upper bound.
\begin{lemma}\label{lem:}For a natural conjunctive query $Q=\exists Y.\ Q'$, $\alpha(Q)\le H(Q)$.
\end{lemma}

\begin{proof}
For a database~$\str D$, 
let $U_{\str D}$ be the random variable with values in $Q'(\str D)$, described as the result $r'$ of the following process: first choose uniformly at random a row $r\in Q(\str D)$, and then, choose uniformly at random a row $r'\in Q'(\str D)$ such that $r'[\var(Q)]=r$. 
The following claim the follows by definition.

\begin{claim}\label{claim:uni}
 The distribution of the random variable $U_\str D[\var(Q)]$ is uniform.
\end{claim}

Now we get that:
\begin{multline*}
\alpha(Q,\mathbb{D})\stackrel{\textit{def}}= \frac{\log|{Q(\mathbb{D})}|}{\max_{R\in Q'}~\log{|R(\mathbb{D}|)}}\stackrel{\textit{Claim \ref{claim:uni}}}{=}\\ =\frac{H(U_{\str D}[\var(Q)])}{\max_{R\in Q'}~\log{|R(\mathbb{D})|}}
 \leq \frac{H(U_{\str D}[\var(Q)])}{\max_{R\in Q'}~ H(U_{\str D}[\var(R)])}\\= H(Q,U_{\str D}).
\end{multline*}
By a similar argument as in the proof of Lemma~\ref{lem:H-alpha}, we derive that $\alpha(Q)\le H(Q)$.
\end{proof}

\subsection{Bag semantics}\label{sec:bag}

In the bag semantics, tuples may occur in the output with multiplicity other than $1$, i.e. if 
$Q$ is a conjunctive query, then $Q_{bs}(\str D)$
is a multiset rather than a set, where the multiplicity of a row in the outcome is equal to the number of rows which are projected to it. This is the standard semantics used in SQL.  The possibility of having high multiplicity of tuples in the output may affect size of the bound of output. Analogously to the definition in the introduction, 
we define the size-increase bound  $\alpha_{bs}(Q)$
for bag semantics as the smallest value $\alpha$
such that $|Q_{bs}(\str D)|\le c\cdot |\str D|^\alpha$
for some constant $c$.

Under bag semantics, the evaluation of a natural join query (without projections) is the same as the evaluation under set semantics. Only projection needs to be handled differently: under bag semantics, tuple repetitions are not removed. This leads us to the following fact: 

\begin{fact}\label{fact:same_value}
 Let $Q=\exists Y~Q'$ be a natural conjunctive query. Then for an arbitrary database $\str D$, the size of the output  $|Q_{bs}(\mathbb{D})|$ under bag semantics is equal to the size of the output $|Q'(\mathbb{D})|$. 
\end{fact}
 
 By the fact above, in order to compute $\alpha_{bs}(Q)$ for the bag semantics we can compute $\alpha_{bs}(Q')$. But $Q'$ is a natural join query, and so, set semantics and bag semantics coincide.  By Theorem \ref{thm:optimization} we get $\alpha(Q')=H(Q')$. Concluding:

\begin{corollary}
 For an arbitrary natural conjunctive query $Q=\exists Y~ Q'$ under bag semantics we have $\alpha_{bs}(Q)=H(Q')$.
\end{corollary}

\section{Evaluation}\label{sec:complexity}
In this section, we give some rudimentary results describing 
bounding the worst-case complexity of computing 
the result of a query $Q(\str D)$, for a given database $\str D$.
In the absence of functional dependencies, it is known~\cite{NgoPRR12} that $Q(\str D)$ can be computed from $\str D$ in time 
proportional to $|\str D|^{\alpha(Q)}$. In the bounds presented below, there is an additional factor $|\str D|^m$, where $m$
is a parameter depending on the functional dependencies, defined below.

Let $\cal F$ be a set of functional dependencies over attributes $X$.
A \emph{minimal component} $C$ is an inclusion-minimal nonempty set of attributes $C\subset X$ with the property that 
whenever $Y\mapsto x$ is a functional dependency
with $Y\cap C$ nonempty, then $x\in C$. Define ${\cal F}[C]$ 
to be the set of functional dependencies over $C$ 
which consists of those functional dependencies $Y\mapsto x$ from $\cal F$,
such that $Y\subseteq C$ (and then necessarily $x\in C$ by definition).

We say that a set of attributes $S\subset X$ \emph{spans} $\cal F$,
if  the smallest subset $\bar S$ of $X$ set containing $S$ and closed under functional dependencies (i.e., $Y\mapsto x$ and $Y\subset \bar S$ implies 
$x\in\bar S$) is equal to $X$.
We say that $\cal F$ has \emph{width} $m$ if it has a spanning set 
of size $m$.
We inductively define the \emph{iterative width} of $\cal F$  to be equal
to $m$, if for every its minimal component $C$, ${\cal F}[C]$ has width  $m$, and after removing from $\cal F$ the attributes which belong to the minimal components, the resulting set of
functional dependencies also has iterative width $m$, or the set
of attributes is empty.

\begin{example}
  If $\cal F$ is an empty set of  functional dependencies 
  over a nonempty set of attributes, then $\cal F$ has 
  iterative width $1$.
\end{example}

\begin{example}
  The set of dependencies $x\mapsto y,y\mapsto z,z\mapsto x$
  has width $1$, since it is spanned by $\set{x}$. It also has 
  iterative width $1$.
\end{example}

\begin{example}
    Let $X$ be a set with three elements and let $\cal F$ be the 
    set of all dependencies of the form $Y\mapsto x$, with $Y\subset X,|Y|=2$. Then $\cal F$ has width $2$.
\end{example}

Let $\str D$ be a database and let $C$ be a minimal component. 
For a row $r$ over attributes $X$, denote by $r/C$ the row with attributes $X$, defined as follows:
$$r/C[x]=\begin{cases}r[x]&\text{ for $x\in X-C$}\\x&\text{ for $x\in X\cap C$}.\end{cases}$$
Therefore, $r/C$ is obtained by replacing each value of an attribute in $C$
by a placeholder, storing the name of the attribute.
Denote by $\str D/C$ the database obtained from $\str D$ by replacing
in each table $R$, every row $r$ by the row $r/C$.
Clearly, we have the following.

\begin{lemma}\label{lem:}
  The database $\str D/C$ can be computed from $\str D$ in linear time.
\end{lemma}

The following lemma is immediate, by the fact that $C$ is a minimal component.

\begin{lemma}\label{lem:}
  The database $\str D/C$ satisfies all the functional dependencies of $Q$.
\end{lemma}

Note, however, that the database $\str D/C$ usually satisfies more functional
dependencies than $\str D$, namely, it satisfies all functional dependencies 
$\emptyset\mapsto x$, for $x\in C$.

\begin{lemma}\label{lem:minimal}Let $C$ be a minimal component, and suppose that
  ${\cal F}[C]$ has width $m$.
  Then, for a given database $\str D$, the result $Q(\str D)$ can be computed from $Q(\str D/C)$ in 
  time $O(|Q(\str D/C)|\cdot |\str D|^m)$.
\end{lemma}

\begin{proof}
  To compute $Q(\str D)$, proceed as follows.
  
  For each row $s\in Q(\str D/C)$, we need to determine whether the placeholders can be replaced by actual values, yielding a row $r$
  such that $r[\var (R)]\in R(\str D)$ for every relation name $R$.
  
  For an arbitrary attribute $x\in C$, consider the set
  $$V_x=\bigcap_{R\in Q}\set{r[x]: r\in R(\str D)}$$ of all possible values for $x$. The set $V_x$ has size at most $|\str D|$, and can be computed
  in linear time from $\str D$.
  
  Let $S$ be a set spanning $C$, 
  Let $K$ denote the table consisting of those rows $r$ over $S$
  such that $r[x]\in V_x$ for all $x\in S$.
  The table $K$ has size at most $|\str D|^m$.

  For a row $s\in Q(\str D/C)$ and a row $r\in K$,
  we say that a row $t$ is \emph{compatible} with $s$ and $r$
  if the following conditions hold:
  \begin{itemize}
    \item    $t[x]=s[x]$ for $x\in S$,
    \item $t[x]=r[x]$ for $x\not\in C$,
    \item $t[\var(R)]\in R(\str D)$ for every relation name $R$.
  \end{itemize}
  
  \begin{claim}
    If there is a row $t$ compatible with $s$ and $r$,
    then it is unique, and, given $s$, can be computed
     in constant time (assuming linear time preprocessing independent of $s$).
  \end{claim}

  The algorithm for computing $Q(\str D)$ proceeds by computing,
  for all $s\in Q(\str D/C)$ and all $r\in K$,
  the row compatible with $s$ and $r$ (if it exists),
  and adding it to the result.
\end{proof}

By iteratively applying Lemma~\ref{lem:minimal},
we obtain the main result of this section.
\begin{prop}\label{pro:complexity}
  Fix a natural join query $Q$
  and functional dependencies $\Ff$ of iterative width $m$. 
   Then there 
  is an algorithm which for a given database $\str D$ computes $Q(\str D)$
  in time $O(|\str D|^{\alpha(Q)+m})$.
\end{prop}

Observe that when the set $\Ff$ of functional dependencies is empty, Proposition~\ref{pro:complexity}
gives the algorithm from~\cite{agm}, whose running time 
is $O(|\str D|^{\alpha(Q)+1})$, since in this case, $\Ff$ has iterative width $1$.

\section{Conclusion and future work}
We characterized the worst-case size-increase 
for the evaluation of conjunctive queries, in two ways:
in terms of entropy and in terms of finite groups. Our characterization improves a construction and answers a question from~\cite{gottlob}.
We also presented a rudimentary result concerning the evaluation of natural join queries.

We see two main directions of a possible future work. One can try to find a method for computing $\alpha(Q)$. This looks hard, and probably will require a deeper understanding of entropy, and the entropy cone in particular. 
By comparison, we note that in cryptography and information theory the following, seemingly similar optimization problem, has emerged~\cite{secret} and is considered notoriously difficult. Given an \emph{access structure}, which is an upward-closed family $A$ of subsets 
of a finite set of {participants} $U$, (i.e., $A\subset P(U)$
and $V\in A,V\subset W\subset U$ imply $W\in A$),
the aim is to find a secret-sharing scheme in which 
a set set of participants $V\subset U$ can jointly determine the secret $s$ if and only if $V\in A$. The goal is to minimize the ratio of information possessed by the participants to the information stored in the secret.
As an optimization problem, this can be expressed as follows.
For $v$ ranging over $\overline{\Gamma_{A\cup\set s}^\ast}$,
satisfying 
$$
\begin{cases}
   v_{X\cup\set{s}}=v_X&\text{if }X\in A,\\
   v_{X\cup\set s}=v_X+v_s&\text{if } X\not\in A,\\
   v_u\le 1&\text{for }u\in U,
\end{cases}
$$
maximize the value $v_s$. The inverse of the optimal value is called the \emph{optimal complexity} or \emph{information rate} of the access scheme $A$.
As noted in~\cite{Farras:2012:OBS:2158958.2158975},
\emph{``determining the optimal complexity for general access structures has appeared to be an extremely difficult open problem''.
}

This  demonstrates that optimization problems over the entropy cone can be very difficult. Of course, this will depend very much on the structure of the problem, which may
in some cases turn out to be feasible. For instance, 
the AGM bound demonstrates that in the absence 
of functional dependencies, it is sufficient to relax 
the entropy cone to Shannon's information inequalities. Whereas this is no longer true in the presence of functional dependencies, as demonstrated in~\cite{gottlob},
it still might be the case that considering only finitely many \emph{non-Shannon} information inequalities is sufficient to compute the optimum.

%
%
%
%
%
%
%
%
%

The paper~\cite{NgoPRR12} manages to find an optimal algorithm for evaluating natural join queries, by proving algorithmic versions of the Loomis-Whitney and Bollob\'as-Thomason inequalities. 
Perhaps it is possible to extend this algorithm to the case with functional dependencies, yielding an optimal running time $|\str D|^{\alpha(Q)}$, even if the precise value of $\alpha(Q)$ is unknown.
 However, the worst-case optimal algorithm presented in \cite{NgoPRR12} uses an optimal solution for fractional edge covering. This suggests that finding a worst-case optimal algorithm for the general case will be impossible without simultaneously computing $\alpha(Q)$.


The fractional edge cover was useful in the analysis of the Hypercube algorithm \cite{BeameKS13}, an algorithm for parallel evaluation of queries. Perhaps some ideas from the current paper can also be applied in the parallel setting.

\bibliographystyle{alpha}
\bibliography{AGM}.

\end{document}